\documentclass[onecolumn,preprintnumbers,elsart]{revtex4}
\usepackage{makeidx}
\usepackage{amssymb}
\usepackage{amsmath}
\usepackage{mathrsfs}
\usepackage{graphicx}
\usepackage{dcolumn}
\usepackage{bm}
\usepackage[center]{subfigure}
\usepackage{color}

\begin{document}

\title{$\mathcal{PT}$-symmetric and antisymmetric nonlinear states in a
split potential box}
\author{Zhaopin Chen$^{1}$}
\author{Yongyao Li$^{2}$}
\author{Boris A. Malomed$^{1,3}$}
\affiliation{$^{1}$Department of Physical Electronics, School of Electrical Engineering,
Faculty of Engineering, Tel Aviv University, Tel Aviv 69978, Israel\\
$^{2}$School of Physics and Optoelectronic Engineering, Foshan University,
Foshan 52800, China\\
$^{3}$ITMO University, St.Petersburg 197101, Russia}

\begin{abstract}
We introduce a one-dimensional $\mathcal{PT}$-symmetric system, which
includes the cubic self-focusing, a double-well potential in the form of an
infinitely deep potential box split in the middle by a delta-functional
barrier of an effective height $\varepsilon $, and constant linear gain and
loss, $\gamma $, in each half-box. The system may be readily realized in
microwave photonics. Using numerical methods, we construct $\mathcal{PT}$%
-symmetric and antisymmetric modes, which represent, respectively, the
system's ground state and first excited state, and identify their stability.
Their instability mainly leads to blowup, except for the case of $%
\varepsilon =0$, when an unstable symmetric mode transforms into a weakly
oscillating breather, and an unstable antisymmetric mode relaxes into a
stable symmetric one. At $\varepsilon >0$, the stability area is much larger
for the $\mathcal{PT}$-antisymmetric state than for its symmetric
counterpart. The stability areas shrink with with increase of the total
power, $P$. In the linear limit, which corresponds to $P\rightarrow 0$, the
stability boundary is found in a analytical form.The stability area of the
antisymmetric state originally expands with the growth of $\gamma $, and
then disappears at a critical value of $\gamma $.
\end{abstract}

\maketitle

\section{Introduction}

Although the quantum theory operates complex wave functions, a fundamental
principle is that eigenvalues of physically relevant quantities must be
real. Normally, this condition is satisfied if the underlying Hamiltonian is
Hermitian \cite{qm}. However, it was discovered that Hamiltonians composed
of Hermitian and anti-Hermitian parts, subject to the constraint of the
parity-time ($\mathcal{PT}$) symmetry, also generate real energy spectra,
provided that the strength of the anti-Hermitian part does not exceed a
certain critical value, above which the $\mathcal{PT}$ symmetry breaks down,
i.e., the energy spectra ceases to be real \cite{bender1}-\cite{review2}.
For one-dimensional single-particle Hamiltonians, which include a complex
potential, $U(x)=V(x)+iW(x)$, whose imaginary part is the anti-Hermitian
term in the Hamiltonian, the $\mathcal{PT}$ symmetry implies that the real
and imaginary parts of the potential are, respectively, even and odd
functions of the coordinate \cite{bender1}:
\begin{equation}
V(x)=V(-x),W(-x)=-W(x).  \label{minus}
\end{equation}%
~

While the concept of $\mathcal{PT}$-symmetric Hamiltonians was not
experimentally realized in the framework of the quantum theory, a
possibility was proposed to emulate the symmetry in optical media with
symmetrically placed gain and loss elements \cite{theo1}-\cite{Kominis},
making use of the commonly known similarity between the Schr\"{o}dinger
equation in quantum mechanics and the equation governing light propagation
in the paraxial approximation. In fact, this setting may be considered as a
specific example of the general class of dissipated structures, the concept
of which was developed by I. Prigogine and his collaborators \cite{Prig}.
This prediction was followed by the experimental implementation of the $%
\mathcal{PT}$ symmetry in various optical waveguides \cite{exp1}-\cite{exp7}
and lasers \cite{exp5,Longhi}, as well as in other photonic settings, such
as metamaterials \cite{exp4}, microcavities \cite{exp6}, optically induced
atomic lattices \cite{exp8}, and exciton-polariton condensates \cite{exci1}-%
\cite{exci3}.

The $\mathcal{PT}$ symmetry can be emulated in other waveguiding settings
too, such as acoustics \cite{acoustics1,acoustics2}, optomechanical systems
\cite{om}, and electronic circuits \cite{electronics}. It was predicted too
in atomic Bose-Einstein condensates (BECs) \cite{Cartarius} and magnetism
\cite{magnetism}. In terms of the theory, $\mathcal{PT}$-symmetric
extensions were also elaborated for Korteweg - de Vries \cite{KdV1,KdV2},
Burgers \cite{Zhenya-Burgers}, and sine-Gordon \cite{Cuevas} equations, as
well as in a model combining the $\mathcal{PT}$ symmetry with emulation of
the spin-orbit coupling in optics \cite{HS}.

While the $\mathcal{PT}$ symmetry is a linear property, it may be combined
with intrinsic nonlinearity of the medium in which the symmetry is
implemented, which is usually as the Kerr self-focusing of optical
materials. Usually, such settings are modelled by nonlinear Schr\"{o}dinger
(NLS) equations with complex potentials subject to constraint (\ref{minus}).
In particular, these models give rise to $\mathcal{PT}$-symmetric solitons,
which were considered in a large number of works \cite{soliton}, \cite%
{Konotop}-\cite{unbreakable} (see recent reviewed in Refs. \cite{review1}
and \cite{PT-review2}), and experimentally demonstrated too \cite{exp7}. A
characteristic feature of $\mathcal{PT}$-symmetric solitons and other
nonlinear modes is that they form continuous families, like in conservative
systems \cite{families}, although the $\mathcal{PT}$-symmetry is realized in
dissipative media. In that sense, $\mathcal{PT}$-symmetric systems represent
an interface between conservative models and traditional dissipative ones,
which normally give rise to isolated solutions in the form of dissipative
solitons, which do not form families \cite{diss1}-\cite{diss3}.

The objective of the present work is to introduce a one-dimensional model
which combines the $\mathcal{PT}$ symmetry and cubic self-focusing at the
most basic level. As concerns the spatially even real part of the potential,
$V(x)$ in Eq. (\ref{minus}), its most fundamental version is represented by
the double-well structure \cite{DWP1}-{DWP4}, \cite{book}. In turn, what may
be considered as, arguably, the most basic form of such a potential in one
dimension is a infinitely deep potential box, split in the middle (at $x=0$)
by an infinitely narrow delta-functional barrier \cite{NatPhot,Elad}. In
this work, we combine the real split-box potential with the simplest
imaginary one, represented by constant gain and loss coefficients in two
half-boxes, at $x>0$ and $x<0$, respectively. Microwave photonics, which may
involve cubic nonlinearity (see, e.g., Ref. \cite{Slavin}), offers the most
straightforward possibility to implement this complex potential, with the
box realized as a waveguide with metallic walls, and the central splitter
induced by a metallic strip partly separating the guiding channel in two
\cite{micro}. The symmetric gain and loss may be realized, in the lossy
material filling the waveguide, by installing an amplifier, with an
appropriate value of the gain, at $x>0$. In principle, the same model may be
implemented in BEC too, assuming that the condensate is loaded into an
appropriately shaped trapping potential, with symmetrically placed
amplifying and lossy elements \cite{Cartarius}, but this may be difficult to
achieve in the real experiment.

The model is formulated in detail in Section II. Then, in Section III, we
report an analytically derived stability boundary for the zero state in the
linearized version of the model, which is a nontrivial finding in the
presence of the complex potential. The main problem, which is addressed in
Section IV, is constructing nonlinear $\mathcal{PT}$-symmetric and
antisymmetric states in this system [alias the ground state (GS), and the
first excited state, respectively]. This is done by means of numerical
methods (the imaginary-time integration for the GS, and the Newton's method
for the antisymmetric modes). Further, we focus on identifying existence and
stability boundaries of these states. In particular, a noteworthy finding is
that the stability area is much larger for the antisymmetric state than for
the symmetric one. The paper is concluded by Section V.

\section{The model}

We consider a 1D model based on the $\mathcal{PT}$-symmetric NLS equation
with the cubic self-focusing nonlinearity term, a double-well potential, in
the form of the infinitely deep box, split in the middle by the
delta-functional barrier, and uniform gain and loss applied in two
half-boxes. The NLS equation is written in the normalized form with zero
boundary conditions at edges of the box, $x=\pm 1/2$:

\begin{gather}
i\frac{\partial \psi }{\partial z}=-\frac{1}{2}\frac{\partial ^{2}\psi }{%
\partial x^{2}}-g\left\vert \psi \right\vert ^{2}\psi +\varepsilon \delta
(x)\psi +i\gamma \sigma (x)\psi ,  \label{psi2} \\
\psi \left( x=\pm \frac{1}{2}\right) =0.  \label{b.c.}
\end{gather}%
Here $z$ and $x$ are the propagation distance and transverse coordinate in
the waveguide, which take values, respectively, $z\geq 0$ and $|x|\,\leq 1/2$
(i.e., the width of the waveguide is scaled to be $1$). Further, $%
\varepsilon $ is the strength of the splitting barrier, and the
self-focusing coefficient is normalized to be $g=1$, except for $g=0$ in the
linearized model. Coefficient $\gamma $ in Eq. (\ref{psi2}) represents the
strength of gain-loss term, with $\sigma (x)$ being an odd function of $x$,
which we here chose as the step profile:

\begin{equation}
\sigma (x)=\mathrm{sgn}(x).  \label{sig}
\end{equation}

Stationary solutions to Eq. (\ref{psi2}) are looked for as
\begin{equation}
\psi (x,z)=\exp \left( ikz\right) u(x),  \label{k}
\end{equation}%
where $k$ is a real propagation constant, and complex function $u$ satisfies
equations

\begin{gather}
-ku+\frac{1}{2}\frac{d^{2}u}{dx^{2}}+g\left\vert u\right\vert
^{2}u=\varepsilon \delta (x)u+i\gamma \mathrm{sgn}(x)u,  \label{StatU} \\
u\left( x=\pm \frac{1}{2}\right) =0.  \label{StatUbc}
\end{gather}%
Stationary states are characterized by the the total power,

\begin{equation}
P=\int_{-1/2}^{+1/2}|\psi (x)|^{2}dx  \label{P}
\end{equation}

To analyze stability of stationary states, we search for perturbed solutions
to Eq. (\ref{psi2}) as

\begin{equation}
\psi (x,z)=e^{ikz}[u(x)+v_{1}(x)e^{-i\lambda z}+v_{2}^{\ast }(x)e^{i\lambda
^{\ast }z}]  \label{PtbSol}
\end{equation}%
where $v_{1}(x)$ and $v_{2}(x)$ are infinitesimal perturbation eigenmodes,
and $\lambda $ is the respective instability growth rate. Linearization
around the stationary solutions leads to the following equation:

\begin{equation}
\left(
\begin{array}{cc}
\hat{F}+i\gamma \sigma (x) & -u^{2} \\
(u^{\ast })^{2} & -\hat{F}+i\gamma \sigma (x)%
\end{array}%
\right) \left(
\begin{array}{c}
v_{1} \\
v_{2}%
\end{array}%
\right) =\lambda \left(
\begin{array}{c}
v_{1} \\
v_{2}%
\end{array}%
\right) ,  \label{eigenvalue}
\end{equation}%
where $\hat{F}=-(1/2)d^{2}/dx^{2}-2\left\vert u\right\vert ^{2}+\epsilon
\delta (x)+k$. This equation, with boundary conditions (\ref{StatUbc}), were
solved numerically. The instability is predicted by the existence of
eigenvalues with $\mathrm{Im}(\lambda )\neq 0$.

\section{Analytical results: stability of the zero background}

The first objective of the analysis is stability of the zero solution in the
framework of the present model, which is a nontrivial issue in the presence
of the complex potential. The corresponding eigenmodes and eigenvalues $k$
should be found from the linearized version of Eqs. (\ref{StatU}):
\begin{equation}
-ku+\frac{1}{2}\frac{d^{2}u}{dx^{2}}=\varepsilon \delta (x)u+i\gamma \mathrm{%
sgn}(x)u,  \label{u}
\end{equation}%
the stability implying that $k$ must be real. $\mathcal{PT}$-symmetric
solutions to Eq. (\ref{u}) are singled out by condition%
\begin{equation}
u(-x)=u^{\ast }(x),  \label{*}
\end{equation}%
with $\ast $ standing for the complex conjugation. Accordingly, the
solutions are looked for as
\begin{equation}
u(x)=v(x)+iw(x),  \label{uvw}
\end{equation}%
with the real and imaginary parts subject to the following constraints:%
\begin{equation}
v(-x)=v(x),~w(-x)=-w(x).  \label{evenodd}
\end{equation}

At $x\neq 0$, where $\delta (x)$ does not appear in Eq. (\ref{u}), one can
eliminate $v$ in favor of $w$ in Eq. (\ref{u}), after the substitution of
expression (\ref{uvw}), and thus derive a single equation for $w(x)$:%
\begin{gather}
v(x)=\frac{\mathrm{sgn}(x)}{\gamma }\left( -kw+\frac{1}{2}\frac{d^{2}w}{%
dx^{2}}\right) ,  \label{vw} \\
\frac{1}{4}\frac{d^{4}w}{dx^{2}}-k\frac{d^{2}w}{dx^{2}}+\left( k^{2}+\gamma
^{2}\right) w=0  \label{w}
\end{gather}%
Fundamental solutions to Eq. (\ref{w}) [which, for the time being, do not
take boundary conditions (\ref{StatUbc}) into regard] are looked for in an
obvious form:%
\begin{equation}
u=\mathrm{const}\cdot \exp \left( Qx\right) ,  \label{exp}
\end{equation}%
\begin{equation}
Q^{2}=2\left( k\pm i\gamma \right) .  \label{Q}
\end{equation}%
Equation (\ref{Q}) yields four roots:%
\begin{eqnarray}
Q &=&\pm Q_{r}\oplus iQ_{i},  \label{+_} \\
Q_{r} &\equiv &\sqrt{\sqrt{k^{2}+\gamma ^{2}}+k},  \label{Qr} \\
Q_{i} &\equiv &\frac{\gamma }{\sqrt{\sqrt{k^{2}+\gamma ^{2}}+k}},  \label{Qi}
\end{eqnarray}%
where $\oplus $ in front of $Q_{i}$ stands for a $\pm $ sign, chosen
independently from $\pm $ in front of $Q_{r}$.

A general ansatz for odd eigenmode $w(x)$, which follows from Eqs. (\ref{exp}%
) and (\ref{+_}), and must satisfy the boundary conditions at edges of the
potential box, is%
\begin{equation}
w(x)=a\sin \left( Q_{i}|x|\right) \sinh \left( Q_{r}x\right) +b\cos \left(
Q_{i}x\right) \sinh \left( Q_{r}x\right) +c\sin \left( Q_{i}x\right) \cosh
\left( Q_{r}x\right) ,  \label{ansatz}
\end{equation}%
where coefficient $a$ may be considered as an arbitrary one. In the first
term, the presence of $|x|$ in $\sin \left( Q_{i}|x|\right) $ implies that
the respective term is an odd function of $x$. A possible additional odd
term, $\sim \mathrm{sgn}(x)\cos \left( Q_{i}x\right) \cosh \left(
Q_{r}x\right) $, is not included in Eq. (\ref{ansatz}), as it contradicts
the continuity of $w(x)$ at $x=0$. Then, the substitution of ansatz (\ref%
{ansatz}) in Eq. (\ref{vw}) yields
\begin{gather}
v(x)=\frac{Q_{i}Q_{r}}{\gamma }\left[ a\cos \left( Q_{i}x\right) \cosh
\left( Q_{r}x\right) -b\,\mathrm{sgn}(x)\sin \left( Q_{i}x\right) \cosh
\left( Q_{r}x\right) \right.  \notag \\
\left. +c\,\mathrm{sgn}(x)\cos \left( Q_{i}x\right) \sinh \left(
Q_{r}x\right) \right] .  \label{vansatz}
\end{gather}%
It is relevant to mention that expression (\ref{vansatz}) yields
\begin{equation}
v(x=0)=a\frac{Q_{i}Q_{r}}{\gamma }.  \label{x=0}
\end{equation}%
This result does not contradict the presence of factor $\mathrm{sgn}(x)$ in
Eq. (\ref{vw}), because term $d^{2}w/dx^{2}$ in the same equation contains
contribution $2Q_{i}Q_{r}a~\mathrm{sgn}(x)$ $\cos \left( Q_{i}|x|\right)
\cosh \left( Q_{r}x\right) $, produced by the second derivative of the first
term in Eq. (\ref{ansatz}), and in the ensuing product of the two factors we
use identity $\left( \mathrm{sgn}(x)\right) ^{2}\equiv 1$.

Next, we take care of boundary conditions (\ref{StatUbc}), i.e., $%
w(x=1/2)=v(x=1/2)=0$, as per Eq. (\ref{uvw}). The substitution of
expressions (\ref{ansatz}) and (\ref{vansatz}) in these conditions yields a
system of linear equations for coefficients $b$ and $c$:
\begin{gather*}
\cos \left( \frac{Q_{i}}{2}\right) \sinh \left( \frac{Q_{r}}{2}\right)
b+\sin \left( \frac{Q_{i}}{2}\right) \cosh \left( \frac{Q_{r}}{2}\right) c \\
=-\sin \left( \frac{Q_{i}}{2}\right) \sinh \left( \frac{Q_{r}}{2}\right) a,
\\
-\sin \left( \frac{Q_{i}}{2}\right) \cosh \left( \frac{Q_{r}}{2}\right)
b+\cos \left( \frac{Q_{i}}{2}\right) \sinh \left( \frac{Q_{r}}{2}\right) c \\
=-\cos \left( \frac{Q_{i}}{2}\right) \cosh \left( \frac{Q_{r}}{2}\right) a.
\end{gather*}%
A solution to these equations is relatively simple:%
\begin{eqnarray}
b &=&\frac{\sin Q_{i}}{\cosh Q_{r}-\cos Q_{i}}a,  \notag \\
c &=&-\frac{\sinh Q_{r}}{\cosh Q_{r}-\cos Q_{i}}a.  \label{bc}
\end{eqnarray}

The remaining condition is the jump of the first derivative of the real
part, $v(x)$, at point $x=0$, induced by the delta-function in Eq. (\ref{u}):%
\begin{equation}
\frac{dv}{dx}|_{x=+0}-\frac{dv}{dx}|_{x=-0}=2\varepsilon v(x=0),
\label{jump}
\end{equation}%
while $v(x)$ must be continuous at $x=0$. The substitution of expressions (%
\ref{vansatz}) and (\ref{x=0}), with coefficients (\ref{bc}), in Eq. (\ref%
{jump}) leads to the final equation which determines the spectrum of
eigenvalues $k$ (arbitrary coefficient $a$ cancels out here):%
\begin{equation}
\varepsilon =-\frac{Q_{i}\sin Q_{i}+Q_{r}\sinh Q_{r}}{\cosh Q_{r}-\cos Q_{i}}%
,  \label{final}
\end{equation}%
which, is, eventually, a relation between the barrier's strength, $%
\varepsilon $, and wavenumber of the perturbation eigenmode which is
generated by Eq. (\ref{u}). It is relevant to mention that Eq. (\ref{final})
is meaningful too for $\varepsilon \leq 0$, which implies placing a narrow
potential well at $x=0$, rather than the barrier.

Note that, in the limit of $\gamma \rightarrow 0$, the expansion of Eqs. (%
\ref{Qr}) and (\ref{Qi}) yields%
\begin{eqnarray}
Q_{r} &\approx &\frac{\gamma }{\sqrt{-2k}}\left( 1-\frac{\gamma ^{2}}{8k^{2}}%
\right) ,  \notag \\
Q_{i} &\approx &\sqrt{-2k}\left( 1+\frac{\gamma ^{2}}{8k^{2}}\right) .
\label{small-gamma}
\end{eqnarray}%
Setting $\gamma =0$, Eq. (\ref{small-gamma}) yields $Q_{r}=0$, $Q_{i}=\sqrt{%
-2k}$, and then Eq. (\ref{final}) amounts to%
\begin{equation}
\varepsilon \tan \left( \sqrt{-\frac{k}{2}}\right) =-\sqrt{-2k},
\label{gamma=0}
\end{equation}%
which is precisely Eq. (12) in Ref. \cite{Elad}, where the conservative
version of the model was considered (in that paper, the notation was $%
k\equiv -\mu $). That equation defines eigenvalues of the propagation
constant for linear modes trapped in the conservative potential box with the
barrier ($\varepsilon >0$) or well ($\varepsilon <0$) placed at the center.

\begin{figure}[tbp]
\centering{\label{fig1a} \includegraphics[scale=0.18]{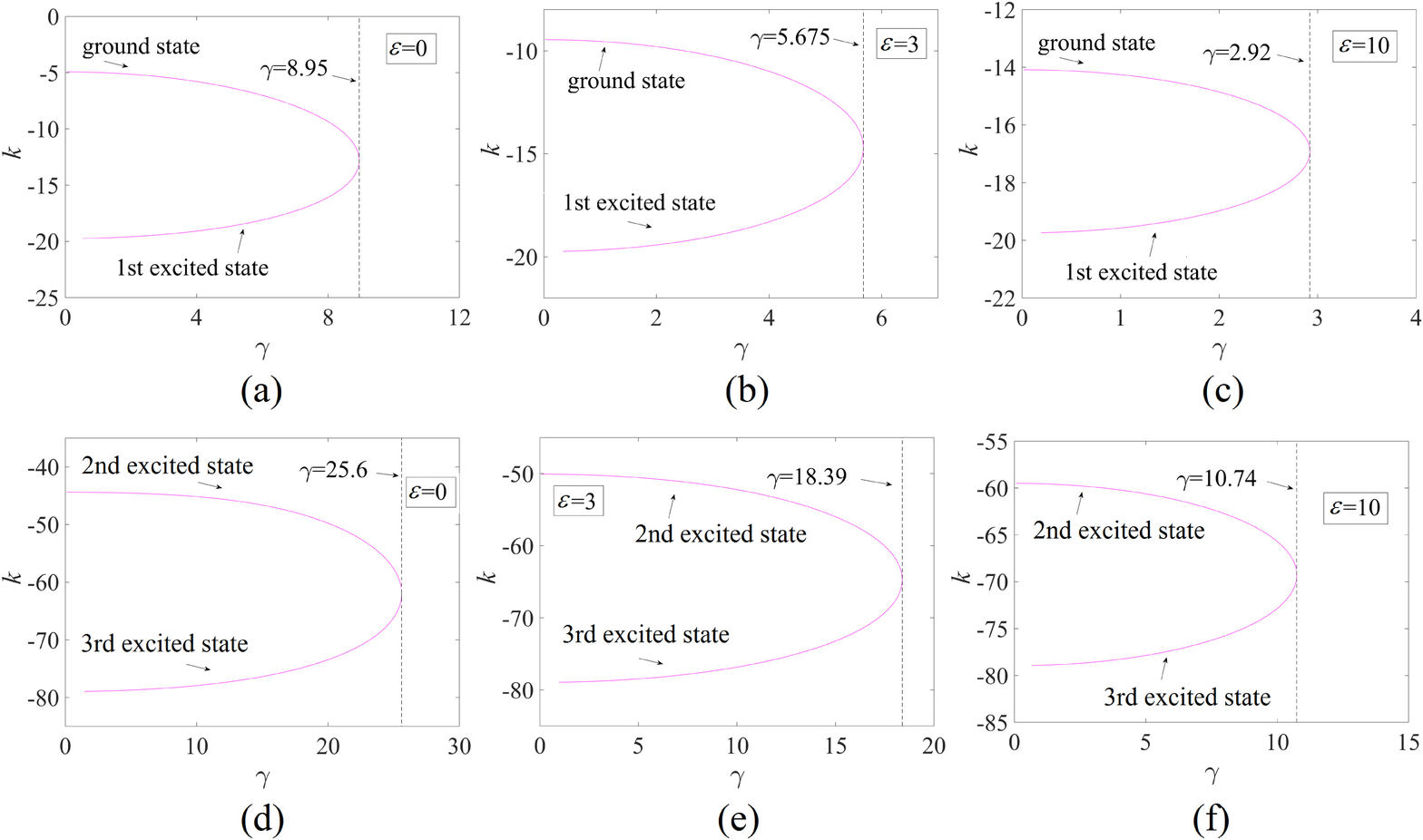}}
\caption{(Color online) Propagation constant $k$ vs. the gain-loss
coefficient, $\protect\gamma $, at fixed values $\protect\varepsilon $,
produced by a numerical solution of Eq. (\protect\ref{final}). Branches
corresponding to the ground state, and the first, second, and third
excited ones, are labeled accordingly.}
\label{linearized1}
\end{figure}

Equation (\ref{final}) was solved numerically, fixing $\varepsilon $ and
gradually increasing $\gamma $, see the result in Fig. \ref{linearized1}. We
aimed to find eigenvalues $k$ for the ground state (GS), along with the first,
second and third excited states, which are identified, respectively,
as the mode corresponding to the smallest value of $|k|$, and subsequent
ones, ordered with the increase of $|k|$. These results clearly show that, for fixed
$\varepsilon$, there is a maximum value, $\gamma _{\max }$, up to which a pair of real
eigenvalues exist, corresponding to the GS and first excited state (panels (a)-(c) in
Fig. \ref{linearized1}). The eigenvalues merge at $\gamma =\gamma _{\max}$, and
become complex at $\gamma >\gamma _{\max}$, which implies breaking of the
$\mathcal{PT}$ symmetry, similar to what
is known in other $\mathcal{PT}$-symmetric systems \cite{bender1}-\cite%
{review2}, except for specific nonlinear ones with unbreakable $\mathcal{PT}$
symmetry \cite{unbreakable}. In addition, the pair including eigenvalues corresponding
to the second and third excited states is displayed too, in panels (d)-(f).

Summarizing these results, in Fig. \ref%
{linearized2} we identify a stability region in the $(\gamma ,\varepsilon )$
plane, where Eq. (\ref{final}) gives rise to pairs of real eigenvalues.
The stability of the zero state requires that all the eigenvalues must be
real, i.e., the stability boundary is given by the lowest curve,
corresponding to the pair of the GS and first excited state.

\begin{figure}[tbp]
\centering{\label{fig2a} \includegraphics[scale=0.25]{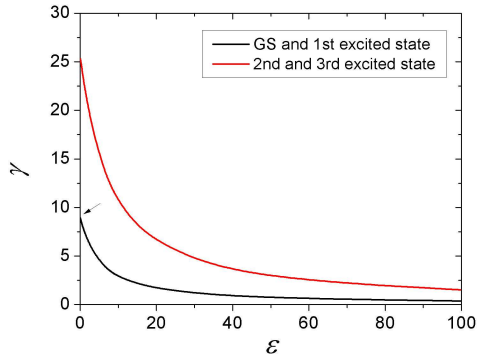}}
\caption{(Color online) Eigenvalues produced by the numerical solution of
Eq. (\protect\ref{final}) for the GS (ground state) and first, second and third
states remain real below the respective boundaries shown in the figure. For
the GS and first state, the boundary starts from point $(\protect\varepsilon ,
\protect\gamma)=(0,8.95)$, as indicated by the black arrow. }
\label{linearized2}
\end{figure}

\section{Numerical results}

\subsection{Symmetric and antisymmetric modes}

Numerical solutions of nonlinear equations (\ref{psi2}), (\ref{b.c.}) and (%
\ref{StatU}), (\ref{StatUbc}) were produced with the ideal $\delta $%
-function replaced by its regularized version,

\begin{equation}
\tilde{\delta}(x)=\frac{1}{\sqrt{\pi }\xi }\exp \left( -\frac{x^{2}}{\xi ^{2}%
}\right) ,  \label{delta}
\end{equation}%
with $\xi \ll 1$. Here, the results are presented for $\xi =0.05$ [taking
smaller $\xi $ does not cause conspicuous differences in the results, except
for Fig. \ref{SyDiag1}(c), see below, where the agreement between the
analytical and numerical stability boundaries improves if smaller $\xi $ is
taken]. Solutions for the GS were generated by applying the imaginary-time
method \cite{ITM1}-\cite{ITM3} to Eqs. (\ref{psi2}) and (\ref{b.c.}). For
producing the first excited state, to which the imaginary-time evolution
cannot converge, the Newton's iteration method was applied directly to Eqs. (%
\ref{StatU}) and (\ref{StatUbc}). The second excited state and higher-order
ones could not be easily found by means of these algorithms. Stability of
the stationary solutions was identified through calculation of the
respective eigenvalue spectra, using Eq. (\ref{eigenvalue}), and then
verified in direct simulations, by means of the Crank-Nicolson scheme.

\begin{figure}[tbp]
\centering{\label{fig3a} \includegraphics[scale=0.16]{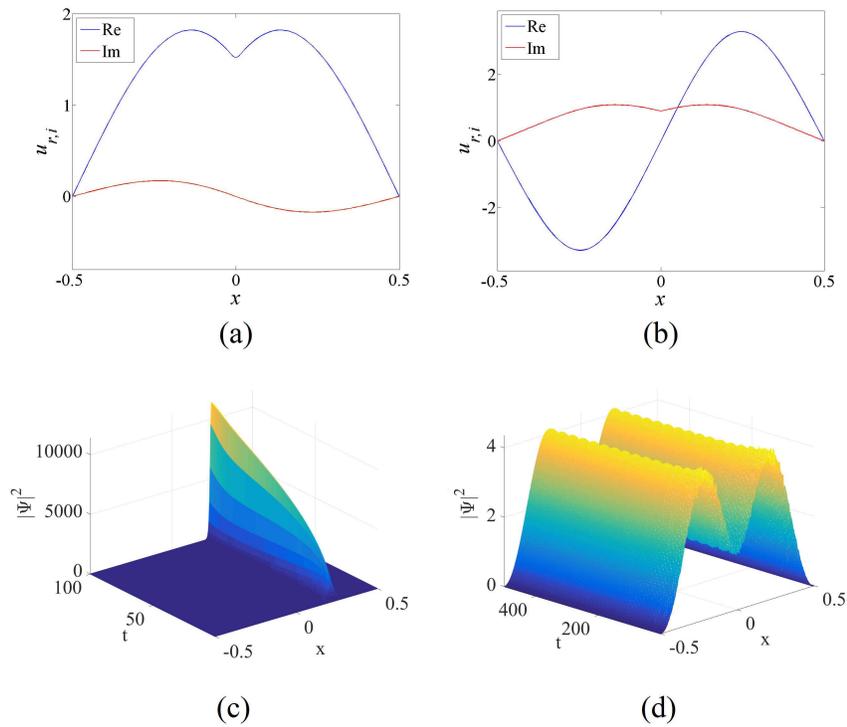}}
\caption{(Color online) Typical examples of a stable $\mathcal{PT}$%
-symmetric mode (a), and antisymmetric one (b), with parameters $(\protect%
\varepsilon ,\protect\gamma ,P,k)=(3,1,2,-7.2150)$ and $(\protect\varepsilon %
,\protect\gamma ,P,k)=(3,3,6.05,-10)$, respectively. (c) The evolution of an
unstable $\mathcal{PT}$-symmetric mode with $(\protect\varepsilon ,\protect%
\gamma ,P,k)=(3,1,5,-3.07)$. (d) The evolution of a weakly unstable $%
\mathcal{PT}$- antisymmetric mode with $(\protect\varepsilon ,\protect\gamma %
,P,k)=(3,5.55,3.0202,-12)$. }
\label{SyAntisyExp}
\end{figure}

Typical examples of stable $\mathcal{PT}$- symmetric and antisymmetric modes
are displayed in Figs. \ref{SyAntisyExp}(a,b). Unstable modes typically
feature exponential growth of perturbations, leading blowup, as shown in
Fig. \ref{SyAntisyExp}(c). Additionally, in very narrow parameter regions
(see Fig. \ref{AntisyStbDiag} below), some antisymmetric modes exhibit weak
oscillatory instability which transforms them into robust breathers via a
supercritical bifurcation (cf. Ref. \cite{Barashenkov}), as shown in Fig. %
\ref{SyAntisyExp}(d). This

\begin{figure}[tbp]
\centering{\label{fig4a} \includegraphics[scale=0.16]{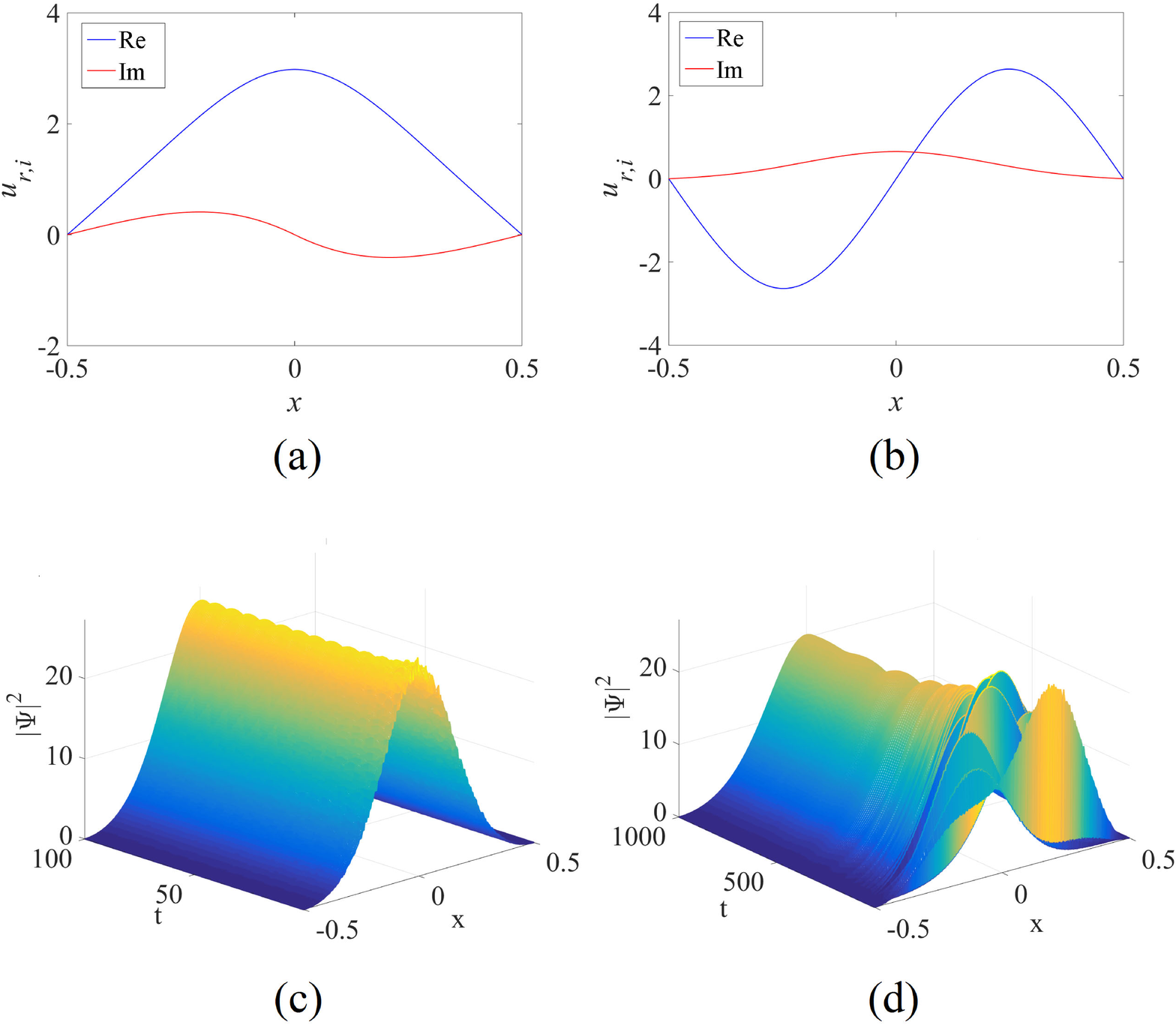}}
\caption{(Color online) The same as in Fig. \protect\ref{SyAntisyExp}, but
for the system without the central barrier ($\protect\varepsilon =0$).
Typical examples of stable $\mathcal{PT}$-symmetric mode (a), and
antisymmetric mode (b), for parameters $(\protect\varepsilon ,\protect\gamma %
,P,k)=(0,3,4,1.0346)$ and $(\protect\varepsilon ,\protect\gamma %
,P,k)=(0,3,3.5593,-14)$, respectively. (c) The evolution of an unstable $%
\mathcal{PT}$-symmetric mode with $(\protect\varepsilon ,\protect\gamma %
,P,k)=(0,10,10,9.333)$. (d) The evolution of an unstable $\mathcal{PT}$%
-antisymmetric mode with $(\protect\varepsilon ,\protect\gamma %
,P,k)=(0,3,8.71,-6)$.}
\label{SyAntisyExp2}
\end{figure}

We have also considered the model without the central barrier, by setting $%
\varepsilon =0$ in Eq. (\ref{psi2}). In this case, it also produces $%
\mathcal{PT}$-symmetric and antisymmetric modes, typical examples of which
are displayed in Figs. \ref{SyAntisyExp2}(a,b). However, if they are
unstable, their instability, shown in Fig. \ref{SyAntisyExp2}(c,d), is
different from what is shown above in Figs. \ref{SyAntisyExp}(c,d). Namely,
unstable $\mathcal{PT}$-symmetric modes transform into breathers, while the
unstable $\mathcal{PT}$-antisymmetric ones transform from the excited state
into the symmetric GS.

\subsection{Existence and stability boundaries for families of the symmetric
and antisymmetric modes}

\begin{figure}[tbp]
\centering{\label{fig5a} \includegraphics[scale=0.15]{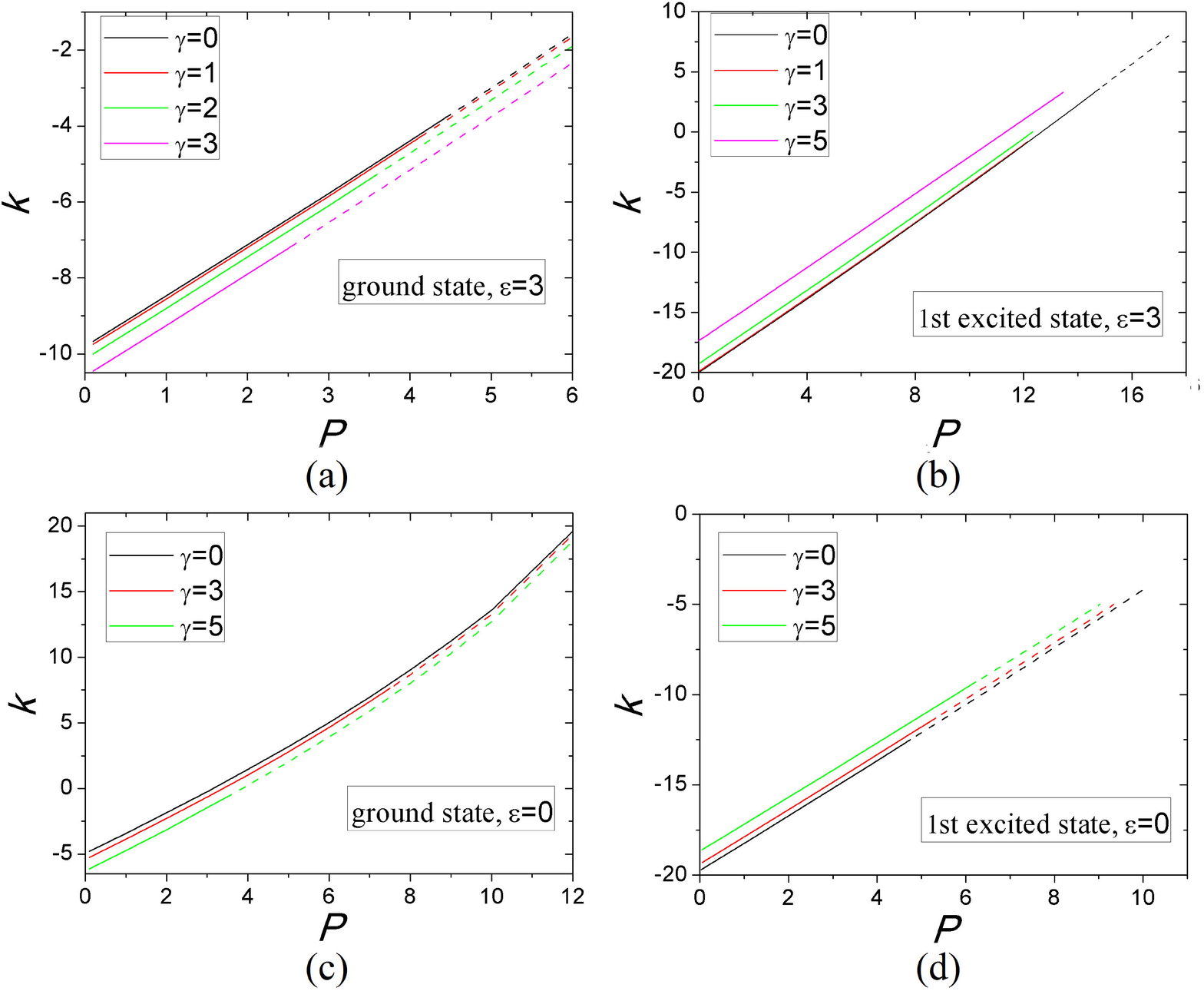}}
\caption{(Color online) Relations between the propagation constant, $k$, and
the total power, $P$, for $\protect\varepsilon =3$ and $0$, at different
fixed values of the gain-loss coefficient, $\protect\gamma $, for the $%
\mathcal{PT}$-symmetric mode (alias the ground state) (a,c) and the $%
\mathcal{PT}$-antisymmetric one (alias the first excited state) (b,d).
Subfamilies of stable and unstable solutions are designated by continuous
and dashed lines, respectively.}
\label{kP}
\end{figure}

Families of $\mathcal{PT}$-symmetric and antisymmetric modes are
characterized by relations between the propagation constant and total power,
$k$ and $P$, which are presented in Fig. \ref{kP}. It is worthy to note that
they all satisfy the known necessary (but not sufficient) Vakhitov-Kolokolov
stability criterion, $dP/dk>0$ \cite{VK1}-\cite{VK3}. We also notice that,
for the GS (symmetric-mode) family, the stability segment shrinks with the
increase of the gain-loss coefficient, $\gamma $. On the other hand, for the
antisymmetric mode, the stability segments becomes shorter as one proceeds
from $\gamma =0$ to $\gamma =1$, but this segment expands with the further
increase of $\gamma $.

\begin{figure}[tbp]
\centering{\label{fig6a} \includegraphics[scale=0.15]{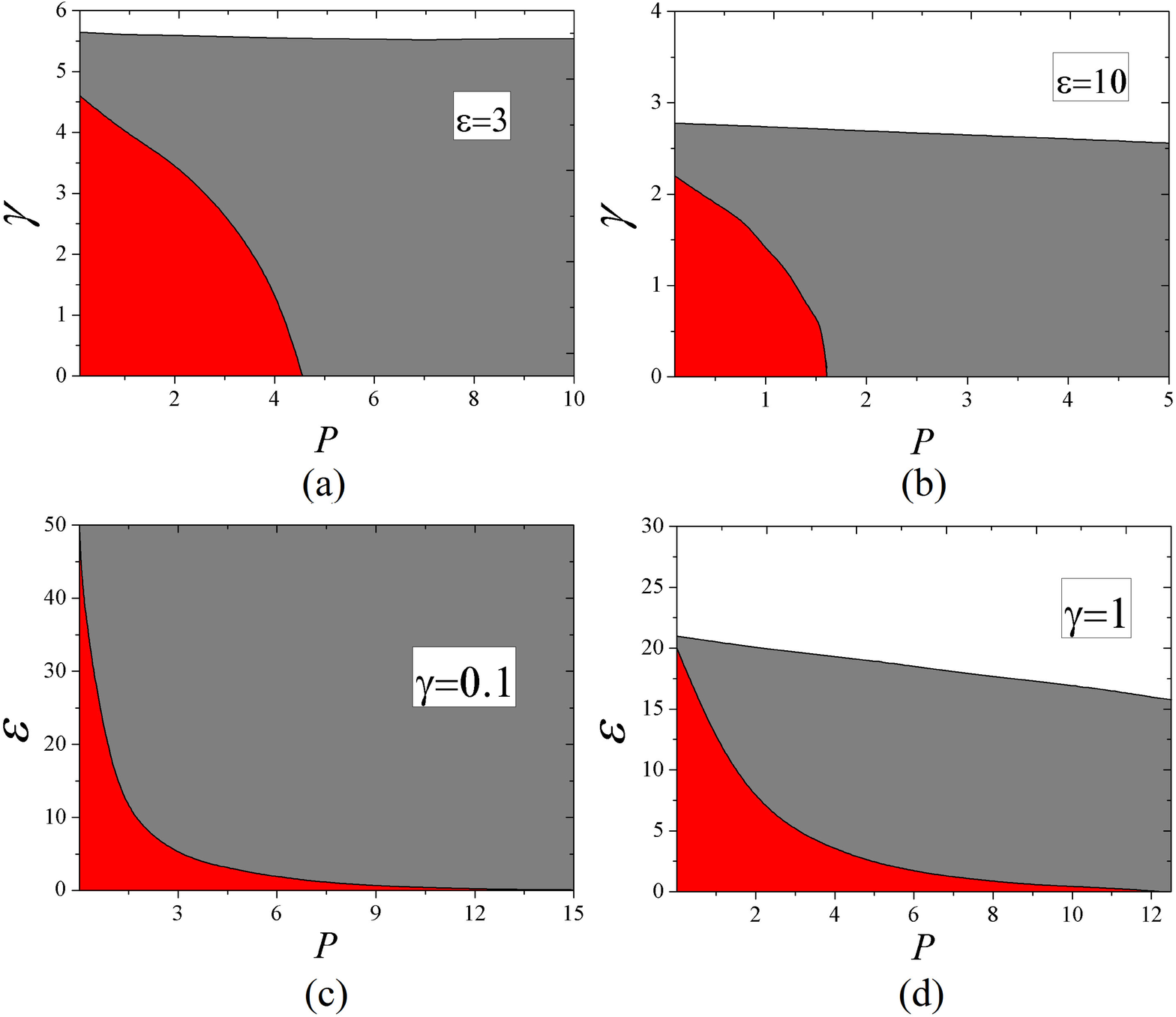}}
\caption{(Color online) Stability diagrams for the $\mathcal{PT}$-symmetric
mode in plane $(\protect\gamma ,P)$ with fixed values of $\protect%
\varepsilon =3$ (a) and $\protect\varepsilon =10$ (b), and in plane $(%
\protect\varepsilon ,P)$, with fixed values of $\protect\gamma =0.1$ (c) and
$\protect\gamma =1$ (d). Stable and unstable states exist in the red and
gray areas, respectively. No solutions were found in white areas.}
\label{SyStbDiag}
\end{figure}

\begin{figure}[tbp]
\centering{\label{fig7a} \includegraphics[scale=0.15]{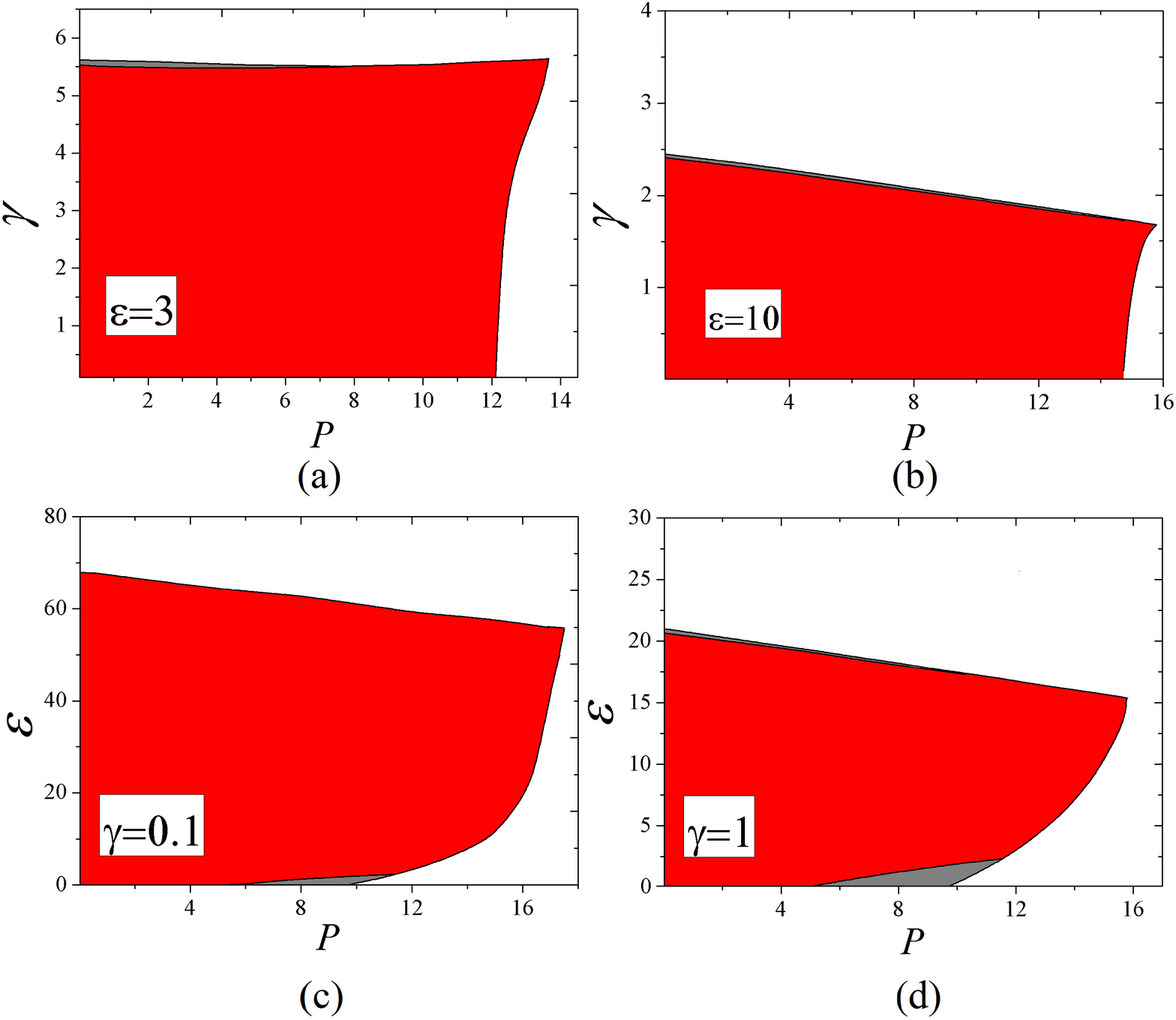}}
\caption{(Color online) The same as in Fig. \protect\ref{SyStbDiag}, but for
the $\mathcal{PT}$- antisymmetric mode. In the top narrow gray areas in
panels (a), (b) and (d) the mode transforms into a weakly oscillating
breather, as shown in Fig. \protect\ref{SyAntisyExp}(d), while unstable
modes suffer blowup in small gray areas at the bottom of panels (c) and (d).}
\label{AntisyStbDiag}
\end{figure}

Results concerning the existence and stability of the symmetric and
antisymmetric \ states are summarized, severally, in parameter planes $%
(\gamma ,P)$ and $(\varepsilon ,P)$ displayed in Figs. \ref{SyStbDiag} and %
\ref{AntisyStbDiag}. First, a salient feature of these results is that
while, at $\varepsilon =0$, the antisymmetric mode has a smaller stability
interval than its symmetric counterpart, its stability area at $\varepsilon
>0$ is dramatically larger than the one for the symmetric mode. This
difference is explained by the fact that the central barrier, imposed by $%
\varepsilon >0$, is favorable for the antisymmetric states, whose wave
function nearly vanishes at $x=0$, and is obviously unfavorable for the
symmetric states, which tend to have a maximum at $x=0$. Eventually, the
antisymmetric states disappear at very larger values of $\varepsilon $,
where the central potential barrier (\ref{delta}), multiplied by a very
large\ $\varepsilon $, suppresses all possible modes in the potential box.

Furthermore, while a trend well-known in many $\mathcal{PT}$-symmetric
systems is that the increase of the gain-loss coefficient, $\gamma $, leads
to the breaking of the $\mathcal{PT}$ symmetry at a critical value of $%
\gamma $ \cite{bender1}-\cite{review2}, the stability region for the
antisymmetric mode in Figs. \ref{SyStbDiag}(a,b) originally demonstrates
slight expansion with the increase of $\gamma $, before the mode disappears
at $\gamma $ exceeding the critical value. On the contrary, the symmetric
mode features the usual trend to the destabilization, following the growth
of $\gamma $, in Figs. \ref{SyStbDiag}(c,d).

Lastly, in the narrow top gray areas in Figs. \ref{AntisyStbDiag}(a,b,d),
the antisymmetric mode is subject to the weak instability shown in Fig. \ref%
{SyAntisyExp}(d), which does not destroy the mode, making it a weakly
oscillating breather. On the other hand, in small gray regions at the bottom
of Figs. \ref{AntisyStbDiag}(c,d), the antisymmetric mode is destroyed by
the blowup instability.

\begin{figure}[tbp]
\centering{\label{fig8a} \includegraphics[scale=0.145]{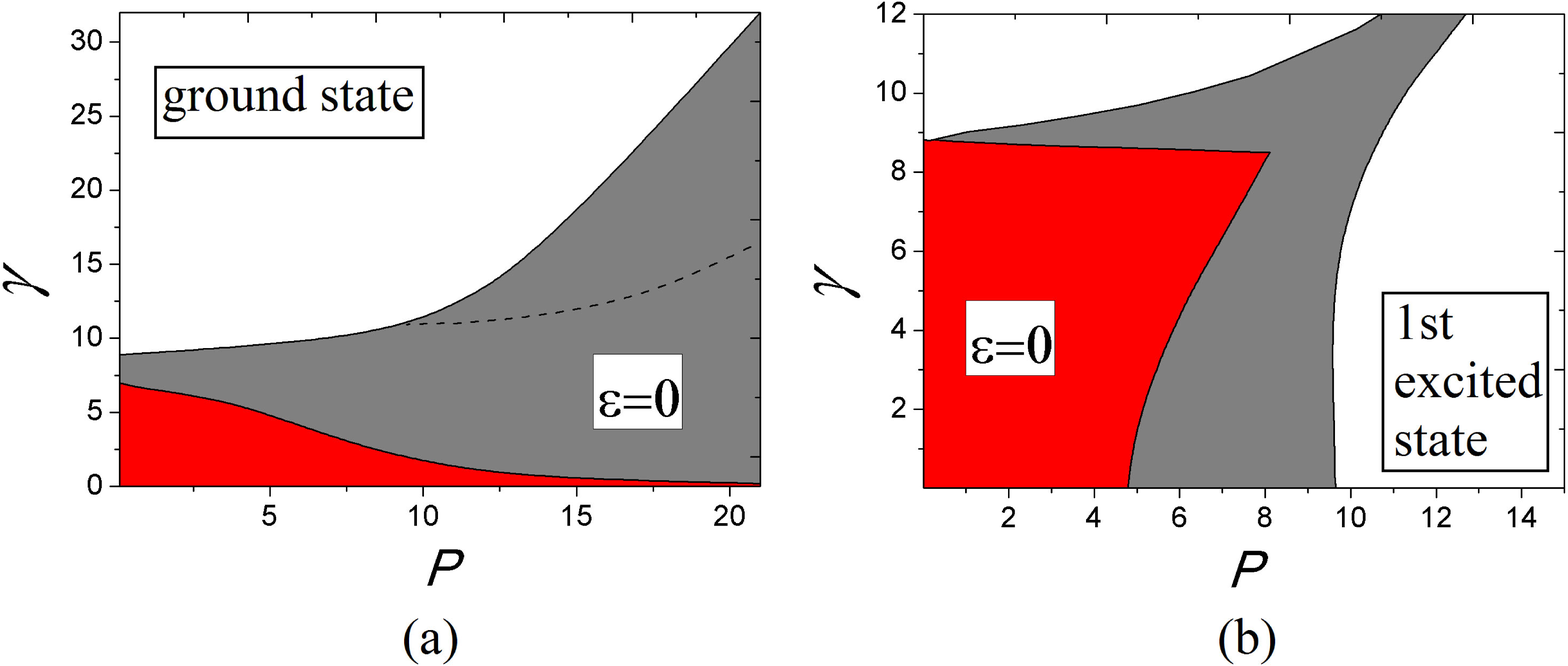}}
\caption{(Color online) Stability diagrams for $\mathcal{PT}$-symmetric (a)
and antisymmetric (b) modes in the system without the box- splitting barrier
($\protect\epsilon =0$). In panel (a), the gray area below the dashed
boundary is populated by weakly unstable symmetric modes, which
spontaneously transform into breathers, as shown in Fig. \protect\ref%
{SyAntisyExp2}(c), while above the boundary the unstable models are
destroyed by the blowup. In panel (b), the instability in the gray area
leads to the spontaneous transformation of unstable $\mathcal{PT}$%
-antisymmetric modes into stable symmetric ones, as shown in Fig. \protect
\ref{SyAntisyExp2}(d).}
\label{SyAntisyStbDiag}
\end{figure}

Stability diagrams for both symmetric and antisymmetric modes in the system
without the splitting barrier (i.e., $\varepsilon =0$) are separately
displayed in Fig. \ref{SyAntisyStbDiag}. Similar to what was noticed above,
the stability region of $\mathcal{PT}$-symmetric mode shrinks with the
increase of $\gamma $ for fixed $\varepsilon $, while, for the antisymmetric
one, it initially expands, and then disappears at a critical value of $%
\gamma $. The unstable symmetric modes feature, respectively, both the
transformation into a weakly oscillating breather [see Fig. \ref%
{SyAntisyExp2}(c)] and the blowup, below and above the dashed boundary in
the gray area in Fig. \ref{SyAntisyStbDiag}(a).

\begin{figure}[tbp]
\centering{\label{fig9a} \includegraphics[scale=0.15]{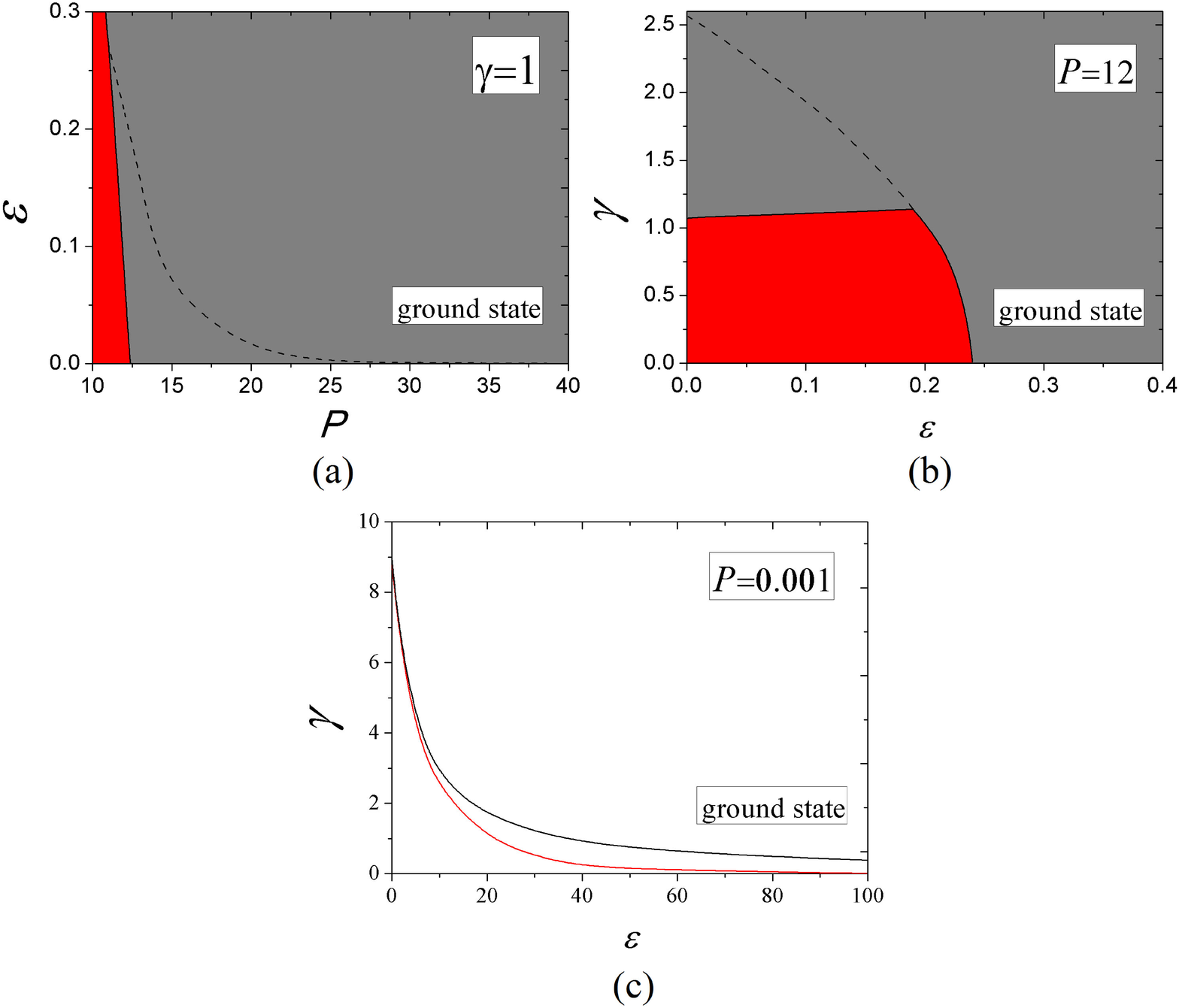}}
\caption{(Color online) Stability diagrams in the $(P,\protect\varepsilon )$
plane with fixed $\protect\gamma =1$ (a), and in the $(\protect\varepsilon ,%
\protect\gamma )$ plane with fixed $P=12$ (b), for $\mathcal{PT}$-symmetric
mode at small values of the barrier's strength, $\protect\varepsilon $.
Regions of the weak oscillatory instability and blowup are located,
severally, below and above dashed boundaries in (a) and (b). (c) The
stability boundaries in the $(\protect\varepsilon ,\protect\gamma )$ plane
for the $\mathcal{PT}$- symmetric modes with a very small total power, $%
P=0.001$, which corresponds to the linearized system. Black and red curves
represent the stability boundaries, as produced, respectively, by the
analytical prediction [see Eq. (\protect\ref{final})] and numerical results.
The symmetric modes are stable beneath the respective boundaries.}
\label{SyDiag1}
\end{figure}

The existence and stability diagrams produced for $\varepsilon =0$ must
continuously extend to $\varepsilon \neq 0$. The continuity is illustrated
by stability diagrams for the GS (symmetric state), plotted for small $%
\varepsilon $ in Fig. \ref{SyDiag1}. In particular, panel (a) demonstrates
that the gray area on the left-hand side of the dashed boundary, populated
by the persistent breathers, shrinks with the increase of $\varepsilon $,
and totally disappear at a small value, $\varepsilon \approx 0.27$. Panel
(b) shows that the same area shrinks with the increase of $\gamma $,
vanishing at $\gamma \approx 2.56$.

In all the cases, the increase of the total power, $P$, leads to
destabilization of the modes, or to their eventual disappearance, as in the
case displayed in Fig. \ref{AntisyStbDiag}. This trend is common for
previously studied nonlinear $\mathcal{PT}$-symmetric systems \cite{bender1}-%
\cite{review2}.

Furthermore, Fig. \ref{SyDiag1}(c) compares the prediction for the stability
boundary of the $\mathcal{PT}$-symmetric mode, as given by the analytically
derived equation (\ref{final}), with the numerically found boundary,
produced by the computation of the stability as per Eq. (\ref{eigenvalue}).
For $\varepsilon \lesssim 10$, the analytical result well matches the
numerical counterpart. At large values of $\varepsilon $, the agreement
breaks down, the numerically identified stability region being much narrower
than predicted analytically. The discrepancy is explained, as mentioned
above in the different context, by the fact that the finite-width potential
barrier (\ref{delta}), multiplied by large $\varepsilon $, strongly changes
the model, in comparison with the underlying one which contains the ideal
delta-function. The discrepancy decreases if smaller $\xi$ is used in Eq. (%
\ref{delta}), but, on the other hand, the use of the splitting barrier with
a finite width corresponds to the physically realistic situation, as the
infinitely narrow barrier cannot be implemented in the experiment.

\begin{figure}[tbp]
\centering{\label{fig10a} \includegraphics[scale=0.15]{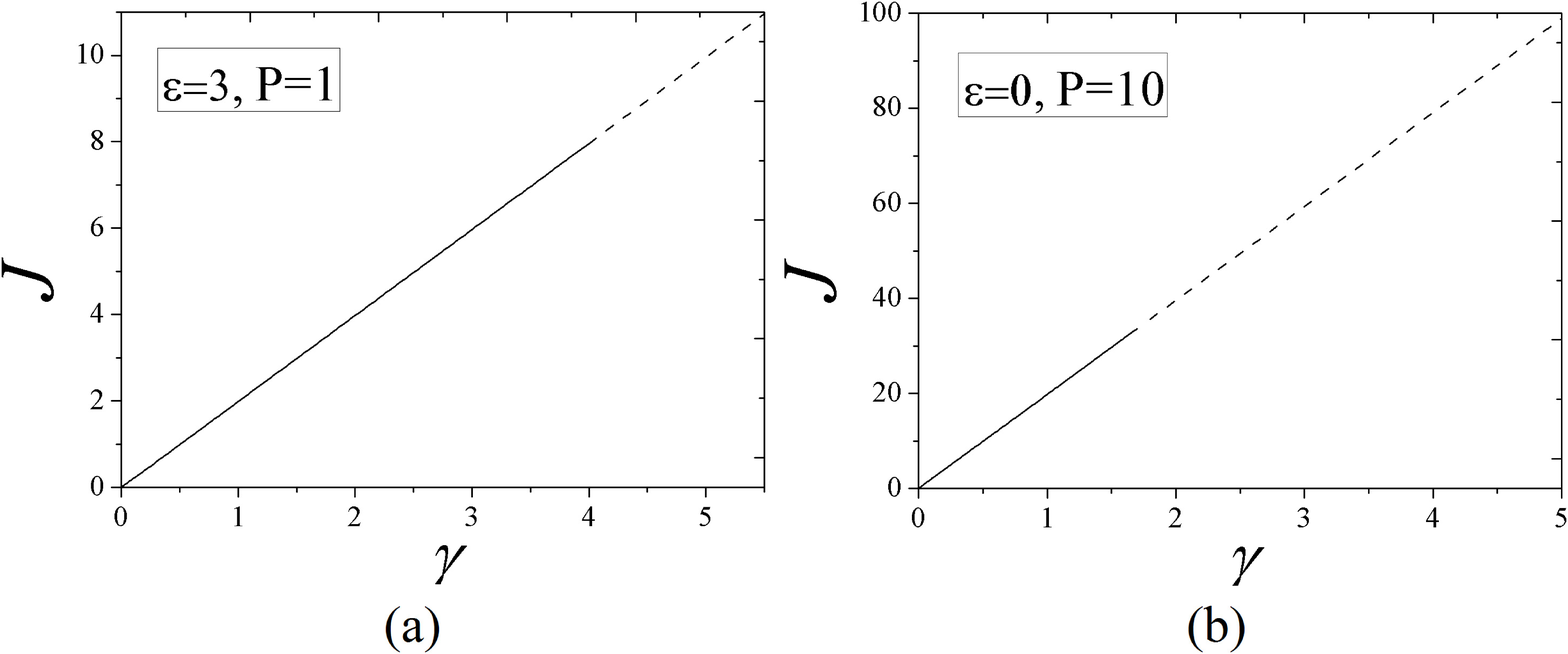}}
\caption{Power flux $J$ across the gain-loss interface,
defined as per Eq. (\protect\ref{flux}), versus the gain-loss coefficient, $%
\protect\gamma $, for the cases of $\protect\varepsilon =3$ (a), and $%
\protect\varepsilon =0$ (b). In each panel, the total power, $P$, is fixed
as indicated. The continuous and dashed segments represent stable and
unstable solutions, respectively.}
\label{flux_g}
\end{figure}

Lastly, $\mathcal{PT}$-symmetric systems are characterized by the flux of
power across the gain-loss interface, defined as
\begin{equation}
J=i\left( u_{x}u^{\ast} - u_{x}^{\ast }u \right) |_{x=0}.  \label{flux}
\end{equation}%
Normally, $J$ is a growing function of the gain-loss coefficient, $\gamma $,
but there are examples of systems demonstrating a \textit{jamming anomaly},
with $dJ/d\gamma <0$. In the present model, typical examples of the $%
J(\gamma )$ dependence are displayed in Fig. \ref{flux_g}. The dependences
are practically linear (the linear form at small $\gamma $ can be easily
explained by the perturbative analysis), without any trace of the jamming
anomaly.

\section{Conclusion}

The objective of this work is to introduce a basic one-dimensional $\mathcal{%
PT}$-symmetric model in the potential box, split into the double-well
potential by the central $\delta $-functional barrier, with strength $%
\varepsilon $. The model includes constant linear gain and loss in two
half-boxes, with strength $\gamma $, which lends the system the $\mathcal{PT}
$ symmetry. The nonlinearity is represented by the usual cubic
self-focusing. The system can be easily realized in guided microwaves and,
in principle, in BEC too.

The stability of the zero state, which is a nontrivial problem for the
present $\mathcal{PT}$-symmetric system, was investigated in the analytical
form. Nonlinear $\mathcal{PT}$-symmetric and antisymmetric modes were found
numerically, using, severally, the imaginary-time-integration and
Newton-iteration methods, and replacing the ideal delta-functional barrier
by a finite-width one. Their stability was explored through numerical
computation of eigenvalues for small perturbations, and verified in direct
simulations. In particular, the analytically predicted stability boundary
for the zero state is confirmed by the numerical results, unless $%
\varepsilon $ is too large. The agreement breaks down at very large values
of $\varepsilon $ because of the difference between the ideal delta-function
and its regularized version used in the numerical calculations. Most of the
unstable modes are destroyed by the blowup, which is typical for $\mathcal{PT%
}$-symmetric systems, but at small values of $\varepsilon $ the symmetric
and antisymmetric modes spontaneously transform, respectively, into weakly
oscillating breathers or stable symmetric GS (ground state). Unstable
antisymmetric states also transform into weakly oscillating breathers in
narrow regions near their existence boundary.

A noteworthy finding is that the stability region at $\varepsilon >0$ for
the antisymmetric (first excited) state is much larger than for the
symmetric GS, which is explained by the fact that the splitting potential
favors antisymmetric configurations, and disfavors symmetric ones. The
stability region of the symmetric states shrinks with the increase of $%
\gamma $ too, while for the the antisymmetric states it originally expands,
but eventually disappears at a critical value of $\gamma $. As usual, the
stability area of all the states shrinks with the increase of their total
power.

A challenging possibility for the extension of the present work is to
develop a two-dimensional counterpart of the model considered here.

\section*{Competing interests}

The authors have no conflict of interests, in the context of this work.

\section*{Authors' contributions}

The model was designed by B.A.M., who was also responsible for the
analytical part of the work. Numerical computations were carried out by Z.C.
and Y.L., who also contributed to the analytical considerations. All the
authors shared the responsibility for drafting the paper.

\section*{Funding}

This work was supported, in part, by grant No. 1286/17 from the Israel
Science Foundation, by grant No. 2015616 from the joint program in physics
between NSF and Binational (US-Israel) Science Foundation, and by NNSFC
(China) through Grant No. 11575063. Z.C. appreciates an excellence
scholarship provided by the Tel Aviv University.

\section*{Acknowledgments}

We appreciate valuable discussions with Nir Dror.


\begin{thebibliography}{99}
\bibitem{qm} L. D. Landau and E. M. Lifshitz, \textit{Quantum Mechanics}
(Nauka Publishers: Moscow, 1974).

\bibitem{bender1} C. M. Bender and S. Boettcher, Real spectra in
non-Hermitian Hamiltonians having $\mathcal{PT}$ symmetry. Phys. Rev. Lett.
\textbf{80}, 5243-5246 (1998).

\bibitem{dorey} P. Dorey, C. Dunning, and R. Tateo, Spectral equivalences,
Bethe ansatz equations, and reality properties in $\mathcal{PT} $-symmetric
quantum mechanics, J. Phys. A: Math. Gen. \textbf{34}, 5679-5704 (2001).

\bibitem{bender2} C. M. Bender, D. C. Brody, and H. F. Jones, Complex
extension of quantum mechanics, Phys. Rev. Lett. \textbf{89}, 270401 (2002).

\bibitem{bender3} C. M. Bender, Making sense of non-Hermitian Hamiltonians,
Rep. Prog. Phys. \textbf{70}, 947-1018 (2007).

\bibitem{bender4} C. M. Bender, Rigorous backbone of $\mathcal{PT}$%
-symmetric quantum mechanics, J. Phys. A: Math. Theor. \textbf{49}, 401002
(2016).

\bibitem{review} K. G. Makris, R. El-Ganainy, D. N. Christodoulides, and Z.
H. Musslimani $\mathcal{PT}$\ symmetric periodic optical potentials, Int. J.
Theor. Phys. \textbf{50}, 1019-1041 (2011).

\bibitem{ptqm} N. Moiseyev, \textit{Non-Hermitian Quantum Mechanics},
(Cambridge University Press, 2011).

\bibitem{review2} L. Feng, R. El-Ganainy, and L. Ge, Non-Hermitian photonics
based on parity-time symmetry, Nature Phot. \textbf{11}, 752-762 (2017).

\bibitem{theo1} A. Ruschhaupt, F. Delgado, and J. G. Muga, Physical
realization of $\mathcal{PT}$-symmetric potential scattering in a planar
slab waveguide, J. Phys. A: Math. Gen. \textbf{38}, L171-L176 (2005).

\bibitem{theo2} R. El-Ganainy, K. G. Makris, D. N. Christodoulides, and Z.
H. Musslimani, Theory of coupled optical $\mathcal{PT}$-symmetric
structures, Opt. Lett. \textbf{32}, 2632-2634 (2007).

\bibitem{soliton} Z. H. Musslimani, K. G. Makris, R. El-Ganainy, and D. N.
Christodoulides, Optical solitons in $\mathcal{PT}$ periodic potentials,
Phys. Rev. Lett. \textbf{100}, 030402 (2008).

\bibitem{theo3} M. V. Berry, Optical lattices with $\mathcal{PT}$-symmetry
are not transparent, J. Phys. A: Math. Theor. \textbf{41}, 244007 (2008).

\bibitem{theo4} S. Klaiman, U. G\"{u}nther, and N. Moiseyev, Visualization
of branch points in $\mathcal{PT}$-Symmetric Waveguides, Phys. Rev. Lett.
\textbf{101}, 080402 (2008).

\bibitem{theo5} S. Longhi, Bloch oscillations in complex crystals with $%
\mathcal{PT}$\ symmetry, Phys. Rev. Lett. \textbf{103}, 123601 (2009).

\bibitem{Konotop} D. A. Zezyulin, Y. V. Kartashov, and V. V. Konotop,
Stability of solitons in $\mathcal{PT}$-symmetric nonlinear potentials, EPL
\textbf{96}, 64003 (2011).

\bibitem{Radik} R. Driben and B. A. Malomed, Stability of solitons in
parity-time-symmetric couplers, Opt. Lett. \textbf{36}, 4323-4325 (2011).

\bibitem{Sukho} N. V. Alexeeva, I. V. Barashenkov, A. A. Sukhorukov, and Y.
S. Kivshar, Optical solitons in $\mathcal{PT}$ -symmetric nonlinear couplers
with gain and loss, Phys. Rev. A \textbf{85}, 063837 (2012).

\bibitem{Bragg} M.-A. Miri, A. B. Aceves, T. Kottos, V. Kovanis, and D. N.
Christodoulides, Bragg solitons in nonlinear $\mathcal{PT}$-symmetric
periodic potentials, Phys. Rev. A \textbf{86}, 033801 (2012).

\bibitem{Yang} S. Nixon, L. Ge, and J. Yang, Stability analysis for solitons
in $\mathcal{PT}$-symmetric optical lattices, Phys. Rev. A \textbf{85},
023822 (2012).

\bibitem{breaking} J. Yang, Symmetry breaking of solitons in one-dimensional
parity-time-symmetric optical potentials, Opt. Lett. \textbf{39}, 5547-5550
(2014).

\bibitem{unbreakable} Y. V. Kartashov, B. A. Malomed, and L. Torner,
Unbreakable $\mathcal{PT}$ symmetry of solitons supported by inhomogeneous
defocusing nonlinearity, Opt. Lett. \textbf{39}, 5641-5644 (2014).

\bibitem{Chen} Z. Chen, J. Liu, S. Fu, Y. Li, and B. A. Malomed, Discrete
solitons and vortices on two-dimensional lattices of $\mathcal{PT}$%
-symmetric couplers, Opt. Express 22, 29679(2014).

\bibitem{Kominis} Y. Kominis, T. Bountis, and S. Flach, Stability through
asymmetry: Modulationally stable nonlinear supermodes of asymmetric
non-Hermitian optical couplers, Phys. Rev. \textbf{95}, 063832 (2017).

\bibitem{Prig} P. Glansdorff and I. Prigogine, \textit{Thermodynamic Theory
of Structures, Stability and Fluctuations} (John Wiley \& Sons: New York,
1971).

\bibitem{exp1} A. Guo, G. J. Salamo, D. Duchesne, R. Morandotti, M.
Volatier-Ravat, V. Aimez, G. A. Siviloglou, and D.. N. Christodoulides,
Observation of $\mathcal{PT}$-symmetry breaking in complex optical
potentials, Phys. Rev. Lett. \textbf{103}, 093902 (2009).

\bibitem{exp2} C. E. R\"{u}ter, K. G. Makris, R. El-Ganainy, D. N.
Christodoulides, M. Segev, and D. Kip, Observation of parity-time symmetry
in optics. \textit{Nature Phys.} \textbf{6}, 192-195 (2010).

\bibitem{exp3} A. Regensburger, C. Bersch, M. A. Miri, G. Onishchukov, D. N.
Christodoulides, and U. Peschel, Parity-time synthetic photonic lattices,
Nature \textbf{488}, 167-171 (2012).

\bibitem{exp7} M. Wimmer, A. Regensburger, M. A. Miri, C. Bersch, D. N.
Christodoulides, and U. Peschel, Observation of optical solitons in $%
\mathcal{PT}$-symmetric lattices, Nature Commun. \textbf{6}, 7782 (2015).

\bibitem{exp5} H. Hodaei, M. A. Miri, M. Heinrich, D. N. Christodoulides,
and M. Khajavikhan, Parity-time-symmetric microring lasers, Science \textbf{%
346}, 975-978 (2014).

\bibitem{Longhi} S. Longhi, $\mathcal{PT}$-symmetric laser absorber, Phys.
Rev. A \textbf{82}, 031801 (2010).

\bibitem{exp4} G. Castaldi, S. Savoia, V. Galdi, A. Al\`{u}, A. and N.
Engheta, $\mathcal{PT}$ metamaterials via complex-coordinate transformation
optics, Phys. Rev. Lett. \textbf{110}, 173901 (2013).

\bibitem{exp6} B. Peng, \c{S}. K. \"{O}zdemir, W. Chen, F. Nori, and L.
Yang, Parity-time-symmetric whispering gallery microcavities, Nature Phys.
\textbf{10}, 394-398 (2014).

\bibitem{exp8} Z. Zhang, Y. Zhang, J. Sheng, L. Yang, M. A. Miri, D. N.
Christodoulides, B. He, Y. Zhang, and M. Xiao, Observation of parity-time
symmetry in optically induced atomic lattices, Phys. Rev. Lett. \textbf{117}%
, 123601 (2016).

\bibitem{exci1} J.-Y. Lien, Y.-N. Chen, N. Ishida, H.-B. Chen, C.-C. Hwang,
and F. Nori, Multistability and condensation of exciton-polaritons below
threshold, Phys. Rev. B \textbf{91}, 024511 (2015).

\bibitem{exci2} T. Gao,E. Estrecho, K. Y. Bliokh, T. C. H. Liew, M. D.
Fraser, S. Brodbeck, M. Kamp, C. Schneider, S. H\"{o}fling, Y. Yamamoto, F.
Nori, Y. S. Kivshar, A. G. Truscott, R. G. Dall, and E. A. Ostrovskaya,
Observation of non-Hermitian degeneracies in a chaotic exciton-polariton
billiard, Nature \textbf{526}, 554-558 (2015).

\bibitem{exci3} I. Yu. Chestnov, S. S. Demirchyan, A. P. Alodjants, Y. G.
Rubo, and A. V. Kavokin, Permanent Rabi oscillations in coupled
exciton-photon systems with $\mathcal{PT}$-symmetry, Sci. Rep. \textbf{6},
19551 (2016).

\bibitem{acoustics1} X. Zhu, H. Ramezani, C. Shi, J. Zhu, and X. Zhang, $%
\mathcal{PT}$-Symmetric Acoustics, Phys. Rev. X \textbf{4} 031042 (2014).

\bibitem{acoustics2} R. Fleury, D. Sounas, and A. Al\'{u}, An invisible
acoustic sensor based on parity-time symmetry, Nature Communications \textbf{%
6}, 5905 (2015).

\bibitem{om} X.-W. Xu, Y.-x. Liu, C.-P. Sun, and Y. Li, Mechanical $\mathcal{%
PT}$ symmetry in coupled optomechanical systems, Phys. Rev. A \textbf{92}
013852 (2015).

\bibitem{electronics} J. Schindler, Z. Lin, J. M. Lee, H. Ramezani, F. M.
Ellis, and T. Kottos, J. Phys. A: Math. Theor. \textbf{45}, 444029 (2012).

\bibitem{Cartarius} L. Schwarz, H. Cartarius, Z. H. Musslimani, J. Main, and
G. Wunner, Vortices in Bose-Einstein condensates with $\mathcal{PT}$%
-symmetric gain and loss, Phys. Rev. \textbf{95}, 053613 (2017).

\bibitem{magnetism} J. M. Lee, T. Kottos, and B. Shapiro, Macroscopic
magnetic structures with balanced gain and loss, Phys. Rev. B \textbf{91},
094416 (2015).

\bibitem{KdV1} C. M. Bender, D. C. Brody, and J.-H. Chen, $\mathcal{PT}$
-symmetric extension of the Korteweg-de Vries equation, J. Phys. A: Math.
Theor. \textbf{40}, F153-F160 (2007).

\bibitem{KdV2} A. Fring, $\mathcal{PT}$ -symmetric deformations of the
Korteweg-de Vries equation, J. Phys. A: Math. Theor. \textbf{40}, 4215-4334
(2007).

\bibitem{Zhenya-Burgers} Z. Y. Yan, Complex $\mathcal{PT}$-symmetric
nonlinear Schr\"{o}dinger equation and Burgers equation, Phil. Trans. Roy.
Soc. A - Math Phys. Eng. Sci. \textbf{371}, 20120059 (2013).

\bibitem{Cuevas} J. Cuevas-Maraver, B. Malomed, and P. Kevrekidis, A $%
\mathcal{PT}$-symmetric dual-core system with the sine-Gordon nonlinearity
and derivative coupling, Symmetry \textbf{8}, 39 (2016).

\bibitem{HS} H. Sakaguchi and B. A. Malomed, One- and two-dimensional
solitons in $\mathcal{PT}$-symmetric systems emulating spin-orbit coupling,
New J. Phys. \textbf{18}, 105005 (2016).

\bibitem{review1} V. V. Konotop, J. Yang, and D. A. Zezyulin, Nonlinear
waves in $\mathcal{PT}$-symmetric systems, Rev. Mod. Phys. \textbf{88},
035002 (2016).

\bibitem{PT-review2} S. V. Suchkov, A. A. Sukhorukov, J. H. Huang, S. V.
Dmitriev, C. Lee, and Y. S. Kivshar, Nonlinear switching and solitons in $%
\mathcal{PT}$-symmetric photonic systems, Laser and Photonics Reviews
\textbf{10}, 177-213 (2016).

\bibitem{families} J. Yang, Necessity of $\mathcal{PT}$ symmetry for soliton
families in one-dimensional complex potentials, Phys. Lett. A, \textbf{378},
367-373 (2014).

\bibitem{diss1} B. A. Malomed, Evolution of nonsoliton and \textquotedblleft
quasiclassical\textquotedblright\ wavetrains in nonlinear Schr\"{o}dinger
and Korteweg - de Vries equations with dissipative perturbations, Physica D
\textbf{29}, 155-172 (1987).

\bibitem{diss2} E. V. Vanin, A. I. Korytin, A. M. Sergeev, D. Anderson, M.
Lisak, and L. V\'{a}zquez, Dissipative optical solitons, Phys. Rev. A
\textbf{49}, 2806-2811 (1994).

\bibitem{diss3} E. N. Tsoy, A. Ankiewicz, and N. Akhmediev, Dynamical models
for dissipative localized waves of the complex Ginzburg-Landau equation,
Phys. Rev. E \textbf{73}, 036621 (2006).

\bibitem{ITM1} M. L. Chiofalo, S. Succi, and M. P. Tosi, Ground state of
trapped interacting Bose-Einstein condensates by an explicit imaginary-time
algorithm, Phys. Rev. E 62, 7438-7444 (2000).

\bibitem{ITM2} W. Bao and Q. Du, Computing the ground state solution of
Bose-Einstein condensates by a normalized gradient flow, SIAM J. Sci.
Comput. 25, 1674-1697 (2004).

\bibitem{ITM3} X. Antoine, W. Bao, and C. Besse, Computational methods for
the dynamics of the nonlinear Schr\"{o}dinger/Gross--Pitaevskii equations,
Comp. Phys. Commun. \textbf{184}, 2621-2633 (2013).

\bibitem{VK1} M. Vakhitov and A. Kolokolov, Stationary solutions of the wave
equation in a medium with nonlinearity saturation, Radiophys. Quantum
Electron. \textbf{16}, 783 (1973).

\bibitem{VK2} L. Berg\'{e}, Wave collapse in physics: Principles and
applications to light and plasma waves, Phys. Rep. \textbf{303}, 259 (1998).

\bibitem{VK3} G. Fibich, \textit{The Nonlinear Schr\"{o}dinger Equation:
Singular Solutions and Optical Collapse} (Springer, Cham, 2015).

\bibitem{DWP1} G. J. Milburn, J. Corney, E. M. Wright, and D. F. Walls,
Quantum dynamics of an atomic Bose-Einstein condensate in a double-well
potential, Phys. Rev. A \textbf{55}, 4318 (1997).

\bibitem{DWP2} A. Smerzi, S. Fantoni, S. Giovanazzi, and S. R. Shenoy,
Quantum coherent atomic tunneling between two trapped Bose-Einstein
condensates, Phys. Rev. Lett. \textbf{79}, 4950 (1997).

\bibitem{DWP3} M. Albiez, R. Gati, J. F\"{o}lling, S. Hunsmann, M.
Cristiani, and M. K. Oberthaler, Direct observation of tunneling and
nonlinear self-trapping in a single bosonic Josephson junction, Phys. Rev.
Lett. \textbf{95}, 010402 (2005).

\bibitem{Barashenkov1} I. V. Barashenkov and D. A. Zezyulin, Localised
nonlinear modes in the $\mathcal{PT}$-symmetric double-delta well
Gross-Pitaevskii equation, in: \textit{Non-Hermitian Hamiltonians in Quantum
Physics, Selected Contributions from the 15th International Conference on
Non-Hermitian Hamiltonians in Quantum Physics} (Palermo, Italy, 18-23 May
2015), pp. 123-142. Editors: F. Bagarello, R. Passante, and Ca. Trapani.

\bibitem{DWP4} K. B. Zegadlo, N. Dror, M. Trippenbach, M. A. Karpierz, and
B. A. Malomed, Spontaneous symmetry breaking of self-trapped and leaky modes
in quasi-double-well potentials, Phys. Rev. A \textbf{93}, 023644 (2016).

\bibitem{book} B. A. Malomed, editor, \textit{Spontaneous Symmetry Breaking,
Self-Trapping, and Josephson Oscillations} (Springer: Berlin, 2013).

\bibitem{NatPhot} B. A. Malomed, Symmetry breaking in laser cavities, Nature
Photon. \textbf{9}, 287 (2015).

\bibitem{Elad} E. Shamriz, N. Dror, and B. A. Malomed, Spontaneous symmetry
breaking in a split potential box, Phys. Rev. E \textbf{94}, 022211 (2016).

\bibitem{micro} D. Deslandes and K. Wu, Integrated microstrip and
rectangular waveguide in planar form, IEEE Microwave Wireless Comp. Lett.
\textbf{11}, 68-70 (2001).

\bibitem{Slavin} A. Slavin and V. Tiberkevich, Nonlinear auto-oscillator
theory of microwave generation by spin-polarized current, IEEE Trans.
Magnetics \textbf{45}, 1875-1918 (2009).

\bibitem{Barashenkov} I. V. Barashenkov, S. V. Suchkov, A. A. Sukhorukov, S.
V. Dmitriev and Y. Kivshar, Breathers in $\mathcal{PT}$-symmetric optical
couplers, Phys. Rev. A \textbf{86}, 053809 (2012).

\bibitem{Barashenkov2} I. V. Barashenkov, D. A. Zezyulin and V. V. Konotop,
Jamming anomaly in $\mathcal{PT}$-symmetric systems, New J. Phys. 18, 075015
(2016).
\end{thebibliography}
\end{document}